# Resistance switching properties of stoichiometric and nitrogen implanted silicon nitride nanolayers on n and p-type Si substrates

A. E. Mavropoulis
Institute of Nanoscience and
Nanotechnology, NCSR "Demokritos",
P.O.Box 60037, GR-153 10 Ag. Paraskevi,
Attica Greece
a.mavropoulis@inn.demokritos.gr

P. Karakolis
Institute of Nanoscience and
Nanotechnology, NCSR "Demokritos",
P.O.Box 60037, GR-153 10 Ag. Paraskevi,
Attica Greece
p.karakolis@inn.demokritos.gr

N. Vasileiadis
Institute of Nanoscience and
Nanotechnology, NCSR "Demokritos",
P.O.Box 60037, GR-153 10 Ag. Paraskevi,
Attica Greece
n.vasiliadis@inn.demokritos.gr

L. Sygellou
Institute of Chemical Engineering Science,
FORTH, GR-26500 Patras, Greece
sygellou@iceht.forth.gr

E. Stavroulakis
Department of Electrical and Computer
Engineering, Democritus University of
Thrace, Xanthi 67100, Greece
emstavro@ee.duth.gr

V. Ioannou-Sougleridis
Institute of Nanoscience and
Nanotechnology, NCSR "Demokritos",
P.O.Box 60037, GR-153 10 Ag. Paraskevi,
Attica Greece
v.ioannou@inn.demokritos.gr

P. Normand
Institute of Nanoscience and
Nanotechnology, NCSR "Demokritos",
P.O.Box 60037, GR-153 10 Ag. Paraskevi,
Attica Greece
p.normand@inn.demokritos.gr

G. Ch. Sirakoulis
Department of Electrical and Computer
Engineering, Democritus University of
Thrace, Xanthi 67100, Greece
gsirak@ee.duth.gr

P. Dimitrakis
Institute of Quantum Computing and
Quantum Technology, NCSR "Demokritos",
P.O.Box 60037, GR-153 10 Ag. Paraskevi,
Attica Greece
p.dimitrakis@inn.demokritos.gr

*Abstract*—**This paper examines the resistive switching characteristics of LPCVD SiNx MNOS ReRAM cells on both heavily doped n- and p-type silicon substrates, focusing on the effects of nitrogen doping. Detailed comparisons of electrical properties through nitrogen implantation reveal variations in trap density and SET-RESET voltages between n and p conductivity Si substrates. Impedance spectroscopy further elucidates the conductive path formation and its resistance.**

## I. Introduction

The limitation of Flash memory in physical scale and the increasing demand for large-data storage has led the scientific community towards the research of alternative technologies. Resistive Random-Access Memories (ReRAMs) present a possible next generation non-volatile memory candidate and have the perspective to be used as Storage Class Memories [1]. They are a noteworthy solution for neuromorphic, in-memory, edge computing [2, 3, 4] and security applications to form true-random number generators [5, 6]. ReRAM functionality is characterized by the change of the dielectric's resistance in the unit cells under voltage bias, which is the result of the formation of a conductive filament. The change of resistance has different origins, which are based on different physical mechanisms depending on the materials used to realize the MIM structure [7, 8, 9]. In Valence Change Memory cells (VCMs), the conductive path between the top and the bottom electrode consists of oxygen vacancies [10, 11]. Properties such as the high scalability (<10nm), speed (<10ns) and endurance (>$10^{12}$ cycles) justify the choice of ReRAM as a viable candidate for replacing the flash memory [1, 12, 13]. Although most of the research investigations on RS phenomena are explained by filamentary concepts, it should be stressed that non-filamentary RS behaviors have been found too [9].

Amorphous Silicon Nitride ($Si_3N_4$) is a widely used material in NVMs, which is currently used in vertical memories [14, 15] and continues to draw interest as a possible resistance switching material for ReRAM applications. It is a material well known for its high strength and toughness, chemical resistance and especially the immunity to humidity and oxygen related effects [16]. Silicon nitride memories are charge-trapping memories and take advantage of the large concentration of intrinsic defects present even at very thin $SiN_x$ layers. This is mainly attributed to the presence of Si dangling bonds, due to N deficiency [17]. Consequently, the stoichiometry of the silicon nitride films strongly affects the energy gap, the static dielectric constant and the refractive index [17, 18]. These characteristics make silicon nitride a very attractive option for ReRAM CMOS technology. Nevertheless, the properties of $SiN_x$ depend on the fabrication method. Chemical vapor deposition (CVD) is the most reliable technique to deposit very thin $SiN_x$ layers with controllable stoichiometry. Low-pressure (LP) and Plasma Enhanced (PE) CVD are the most mature techniques [19]. Stoichiometric material ([N]/[Si]=x=1.33) can be achieved by LPCVD, while x can be tuned by changing the ratio of the precursors and the deposition conditions [20]. S. Kim *et al.* [21]

compared the behavior of a PECVD Si$_3$N$_4$ to LPCVD Si$_3$N$_4$ ReRAM cells and examined the role of the presence of different kinds of defects in RS. In that report, the proposed switching mechanism that dominates the resistive switching phenomena is the Space Charge Limited Conduction (SCLC). Other studies report a filament switch mechanism based on nitrogen vacancies for silicon nitride ReRAM cells. We have demonstrated in our previous work that the filament is formed from nitrogen vacancies, as well as the fact that we can change the conduction mechanism to Poole-Frenkel by doping the material with Si atoms [22]. Finally, an improvement of the resistive switching characteristics of a PE-CVD silicon nitride ReRAM cell by tuning the concentration of the nitrogen has been proposed [20].

In the present work, the switching behavior of LPCVD SiN$_x$ ReRAM cells utilizing an almost stoichiometric layer is examined and then it is compared to the behavior of a non- stoichiometric nitrogen-implanted SiN$_x$.

## II. EXPERIMENTAL

### A. Sample Fabrication

Both n$^+$ and p$^+$-type 100mm (100) silicon wafers were used as a substrate, named SN5, SNI5 and SP5, SPI5 respectively, with resistivity $\rho < 0.005$ Ω cm. A 2 nm thick tunnel barrier SiO$_2$ layer was thermally grown by dry oxidation in standard horizontal atmospheric pressure furnace at 850°C in O-diluted (10% O$_2$ v/v) ambient. Next, a 5.5 nm LPCVD Si$_3$N$_4$ layer was deposited as mentioned in our previous work [23]. After that, a nitrogen implantation step was performed to dope the SiN$_x$ layer on the SNI5 and SPI5 wafers. The implant energy was set to 25 keV, and the nitrogen implant dose was $1\times10^{13}$ ions /cm$^2$. A 30 nm Cu / 30nm Pt TE bilayer was deposited by sputtering ($5\times10^{-8}$ Torr base pressure, 30W DC Magnetron, 0.2 nm/s deposition rate). The Pt metal layer was used to prevent the oxidation of the copper electrodes. Following the TE formation, a thick 500 nm Al metal layer was deposited on the backside of the wafers, to serve as a back contact metal.

## III. RESULTS AND DISCUSSION

### A. Structural Characterization

A series of characterization tests are performed in order to understand the structure of the fabricated SiN$_x$.

*1) High Resolution Transmission Electron Microscopy (HR-TEM) Investigations*

The fabricated samples were investigated by HR-TEM in cross-section mode. Figure 1a presents the high-resolution cross-sectional TEM micro images of MNOS structure before ion implantation. Evidently, in the MNOS the nitride and oxide layers are 5.5nm and 2nm thick respectively. Silicon nitride layers remain amorphous after the implantation and no swelling of the implanted layers was observed due to the introduction of nitrogen species (Figure 1b). Nevertheless, in the case of the MNOS structures, the 2nm SiO$_2$ is not clearly distinguished from the nitride superficial layer attributed to the severe nitridation process that took place and its possible transformation into oxynitride. The interface of the dielectric with Si substrate is atomically sharp without undulations denoting that the implantation process does not affect the interface quality. Furthermore, no extended defects into the Si substrates are observed due to the implanted nitrogen atoms.

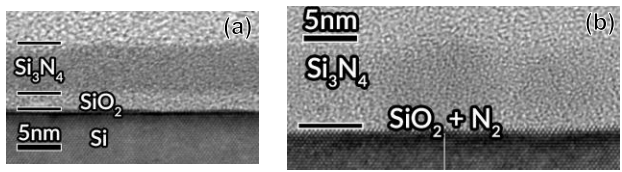

Figure 1 HR-XTEM micro images showing the (a) as deposited and (b) the implanted layers of the fabricated MNOS structures.

### B. Electrical Characterization

The DC current – voltage characteristics (I-V) of the fabricated ReRAM cells were measured using the HP4155A Semiconductor Parameter Analyzer and a manual wafer prober. Impedance Spectroscopy (IS) measurements in the range 100Hz – 1MHz were also performed on devices either in HRS or LRS using the HP4284A Precision LCR Meter. All measurements were performed at room temperature in ambient air (RH=45%).

*1) Current - Voltage characteristics*

Typical I-V characteristics measured for the stoichiometric samples are presented in Figure 2a after multiple sequential SET/RESET voltage sweep cycles. The devices exhibit a typical bipolar behavior, where the transition from the HRS to LRS is achieved by applying a positive voltage bias to the top Cu electrode around +3.5V and the reverse transition (from LRS to HRS) occurs at negative voltages ranging from -2V to -4V. Both transitions are characterized by a sudden "current jump" and "current drop", in positive and negative voltage branches of the I-V curves respectively.

It should be noted that the initial state (fresh state) of every device tested was the HRS and no "forming" step was required. A large memory window, given by log(I$_{ON}$ / I$_{OFF}$), such as $\geq 3$ orders of magnitude was measured at 0.1V, where I$_{ON}$ and I$_{OFF}$ correspond to $I_{LRS}$ and $I_{HRS}$ respectively. Large memory windows are mainly attributed to the insulating properties of the dielectric.

The current compliance ($I_{CC}$) value is critical in voltage sweeps because it governs the cell's resistance at both SET and RESET states. In the SET operation, the $I_{CC}$ controls the creation of the conductive path (filament). If the compliance current used during SET is very high, the creation of many conductive paths and/or thicker conductive filament are possible. In this case, the filaments are not easily destroyed during the RESET operation. Similarly, if $I_{CC}$ is too low, it cannot induce the effective resistance change during the SET process [24], thus incomplete or unstable filaments are created. As a result, the LRS state is not steady. Different $I_{CC}$ values are tested to obtain the lowest current operation possible. In our case, the 100μA was the minimum value required for the SET operation; smaller currents lead to an unstable resistance switch of the cells. In similar fashion, the lowest $I_{CC}$ value achieved for the RESET operation was 5mA. Any value less than 5mA was not sufficient to revert the device back to HRS.

Figure 2c reveals that SP5 ($p^+$-Si substrate) requires higher voltages for the SET operation than the SN samples ($n^+$-Si substrate). More specifically, the $V_{SET}$ value for SP samples is found to be around +4V compared to +3.5V for the SN samples. These results are in accordance with previous research [25]. Also, it is evident that the $I_{CC}$ (5mA) during the RESET operation is again much higher than in the SET operation (100μA). The different $I_{CC}$ values in SET and RESET operations have been reported by many researchers [24] and are mainly attributed to the current overshoot during the SET process because of the faster switching time of the device compared to the response time of the measurement instrument. The current overshoot effect along with the stochastic nature of the conductive filament and the resistance switching in $SiN_x$ of the fabricated devices, causes a large variation of the LRS as well as HRS. Such variability issues have also been observed and reported in other ReRAM devices [14]. The reduction of the current overshoot issue is possible by the careful selection of cell's capacitance and/or the use of a transistor; also, the $I_{CC}$ could efficiently control the RESET voltage and the RESET current [15, 16].

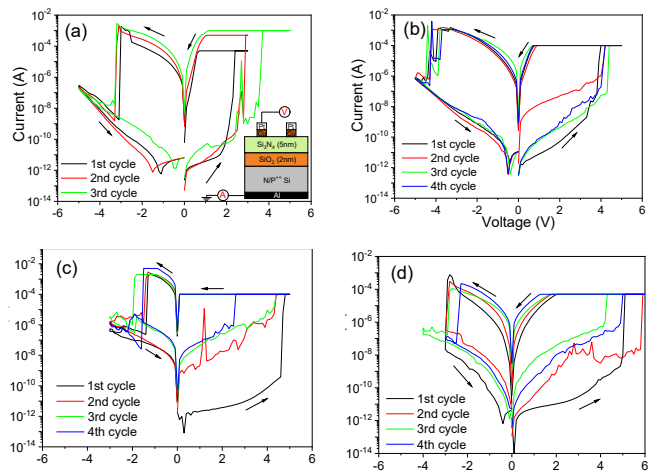

*Figure 2 Experimental I-V characteristics for samples (a) SN5, (b) SNI5, (c) SP5 and (d) SPI5.*

Similar behavior can be observed for the nitrogen-rich samples. In Figure 2b four sequential SET/RESET voltage sweep cycles are presented. The examined cells exhibit bipolar behavior, where the transition from the HRS to LRS is achieved by applying positive voltage bias on TE in the range of +4 to +5V and the reverse transition (from LRS to HRS) occurs at negative voltages of around -4V. These voltages are higher than the ones for the corresponding stoichiometric samples. The fresh state of the examined devices was the HRS and no special "forming" step was followed, just like with the stoichiometric. The log($I_{ON}$ / $I_{OFF}$) ratio is ≥ 3 orders of magnitude. Different $I_{CC}$ values are tested to obtain the lowest current operation possible. The reported minimum $I_{CC}$ for the SET and RESET operation was the same as the stoichiometric samples (100μA and 5mA respectively).

A careful observation of the I-V measurements in Figure 2d reveals that sample SPI5 (Pt/Cu/$SiN_x$/$SiO_2$/$p^+$-Si) requires larger voltages for the SET operation than the other samples. Specifically, the $V_{SET}$ value for the SPI5 sample is found to be around +5V compared to +4V for the other nitrogen rich sample, SNI5. The measured $V_{SET}$ for the SPI is higher than the SP. This is probably attributed to the higher contact resistance of $p^+$-Si substrates compared to $n^+$-Si. Moreover, $V_{RESET}$ for both SP5 and SPI5 is considerably lower than SN5 and SNI5.

Almost in every sample the HRS current increases after each cycle. This is attributed to defects that are permanently created in the dielectric and don't dissolve during the RESET. This can potentially change the read current at each cycle.

*a) Trap density*

From the analysis of the SCLC graphs as mentioned in [23] it is possible to calculate the trap density ($N_t$) of the dielectric materials, using the transition $V_{TR}$ and trap-filled limited $V_{TFL}$ voltages.

Table 1 Trap density of the stoichiometric and doped nitrides.

| Sample | $V_{TR}$ (V) | $V_{TFL}$ (V) | | $N_t$ ($10^{19}$cm$^{-3}$) | | θ | $t_c$ (μs) | $t_d$ (μs) |
|---|---|---|---|---|---|---|---|---|
| | | 1st cycle | 2nd cycle | 1st cycle | 2nd cycle | | | |
| SN5 | 1.4 | 2.1 | 1.4 | 2.60 | 1.74 | 12.60 | 0.91 | 0.30 |
| SNI5 | 1.2 | 1.8 | 1.9 | 2.23 | 2.36 | 14.70 | 1.06 | 0.35 |
| SP5 | 2.2 | 2.9 | 2.5 | 3.60 | 3.10 | 8.02 | 0.58 | 0.22 |
| SPI5 | 1.9 | 2.6 | 3.2 | 3.22 | 3.97 | 9.28 | 0.67 | 0.24 |

It can be seen from Table 1 that the trap density for the stoichiometric samples was higher than that of the nitrogen rich, because the implanted nitrogen atoms passivate part of the Si dangling bonds in the nitride layer. Moreover, the trap density during the 2nd cycle of operation is generally higher. This can be attributed to the formation of the nitrogen vacancies from the conductive filament, which is not completely dissolved during the RESET operation.

*2) Operation parameters statistics*

Figure 3 represents the $V_{SET}$ and $V_{RESET}$ distributions for all the devices tested. The $V_{SET}$ and $V_{RESET}$ statistics confirm the findings derived from the I-V characteristics from the previous section. The stoichiometric samples are characterized by lower $V_{SET}$ and $V_{RESET}$ values (+2.88V, -3.39V for SN5 and +4.20V, -1.41V for SP5) in comparison with the according nitrogen rich (+3.63V, -3.53V for SNI5 and +5.02V, -2.42V for SPI5). As mentioned in Table 1 the trap density is lower for the N-implanted $SiN_x$, this means that there are less Nitrogen sites leading to the formation of N-vacancies and subsequently to form the conductive filament, resulting in higher SET and RESET voltages. Studies have shown that nitrogen-rich $SiN_x$ films typically exhibit lower trap densities compared to stoichiometric $SiN_x$. This is because the additional nitrogen atoms can suppress silicon dangling bonds, leading to fewer defect sites available for conductive filament formation [26]. By calculating the standard deviation to the mean value ratio, we can study the variability in the SET and RESET voltages. The presence of a $SiO_2$ layer creates a better interface between the Si substrate and the $SiN_x$ dielectric. So, it can be assumed that the interface is responsible for that variability.

The variation of the SET voltages is quite similar in all samples. It increases however for the $V_{RESET}$. In addition, the variability is smaller for devices on n-type wafers.

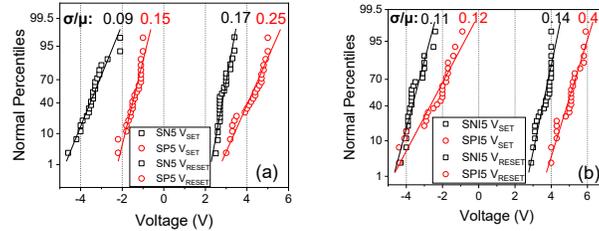

Figure 3 Statistical plots of $V_{SET}$ and $V_{RESET}$ for (a) SN5, SP5 and (b) SNI5, SPI5.

*3) Impedance spectroscopy*

The theory behind the impedance spectroscopy measurements and the measurement methodology have been analyzed previously [22]. The fitting curves are presented, while the fitting parameters of the equivalent circuit ($R_s$ + ($R_p$ ∥ $C_p$) as shown in Figure 4b) are listed in Table 2.

Table 2 Fitting parameters of equivalent circuit for all samples measured at 0.2V bias.

| Sample | $R_s$ (Ω) | $R_p$ (kΩ) | $C_p$ (pF) |
|---|---|---|---|
| SN5 | 83.9 | 220.9 | 55.7 |
| SNI5 | 484.5 | 126.6 | 66.8 |
| SP5 | 284.5 | 77.2 | 54.1 |
| SPI5 | 284.5 | 40.7 | 52.7 |

The resistance $R_p$ is directly related to the resistance of the conductive filament that is formed in the dielectric, and the parallel capacitance $C_p$ is the geometric capacitance of the unswitched $SiN_x$ layer.

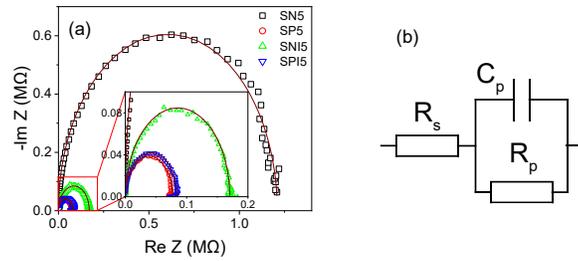

*Figure 4 a) Nyquist plots for samples SN5, SNI5, SP5 and SPI5 at LRS with +0.2V dc bias. Solid lines represent the fitting of the Randles circuit model. b) Equivalent circuit for the Nyquist plots*

The semicircle form of the Nyquist plots, combined with the fitting lines according to Randles circuit, shows that the LRS is governed both by resistive and capacitive effects. Based on calculations derived from the graphs, for the stoichiometric samples $R_s$ values are equal to few hundred Ohm, $R_p$ to few hundred kΩ and $C_p$ values range between 45-65 pF respectively. At low frequencies, the capacitive effect is negligible and both resistors contribute to the conduction ($R_s+R_p$). Since $R_p$ is bigger than $R_s$ by three orders of magnitude, the kinetically slow charge transfer reaction (large value of $R_p$) is the dominant effect. On the other hand, at very high frequencies $R_p=0$ and the resistance of the bulk conducting region is the only factor considered. At the medium frequency range, the capacitive effect becomes significant. On the other hand, for the nitrogen rich samples the $R_p$ was generally lower than the one of the respective stoichiometric nitrides, which indicates that the conductive filament formed is of lower resistance. This is mainly attributed to the lower number of available Si dangling bond defects. It should be stressed that N vacancies are moving via the defects presented in the nitride bulk. Consequently, $C_p$ is slightly higher, ranging from 50pF to 65pF due to the larger unswitched volume of silicon nitride.

## IV. CONCLUSION

In conclusion, this study has thoroughly examined the resistive switching characteristics of LPCVD $SiN_x$ MNOS ReRAM cells on both heavily doped n- and p-type silicon substrates, with a particular focus on the effects of nitrogen doping. The findings indicate that nitrogen-rich samples exhibit higher SET-RESET voltages compared to stoichiometric ones, accompanied by greater variability in these voltages. Impedance spectroscopy and I-V measurements further reveal that stoichiometric samples possess higher SET resistance, impacting the switching behaviour and stability. These insights underscore the potential of silicon nitride for ReRAM applications, emphasizing its adaptability and efficiency in advanced memory architectures. The results contribute to a deeper understanding of the material properties and operational parameters crucial for optimizing ReRAM performance, paving the way for more efficient and scalable memory technologies. To enhance the performance and reliability of silicon nitride-based ReRAM devices, it is necessary to explore the influence of various doping levels and optimize implantation conditions.

ACKNOWLEDGMENT

This work was supported in part by the research project "LIMA-chip" (Proj.No. 2748) which are funded by the Hellenic Foundation of Research and Innovation (HFRI) respectively.